\documentclass{sig-alt-release2}
\usepackage{amsmath}
\usepackage{subfig}
\usepackage{graphicx}
\usepackage{array}

\begin{document}

\conferenceinfo{MobiHeld'09,} {August 17, 2009, Barcelona, Spain.} 
\CopyrightYear{2009}
\crdata{978-1-60558-444-7/09/08} 

\title{Tuning Message Size in Opportunistic Mobile Networks}

\numberofauthors{3}

\author{
\alignauthor
John Whitbeck \\
  \affaddr{Thales Communications and UPMC Paris Universitas}
\alignauthor
Vania Conan \\
  \affaddr{Thales Communications}
\alignauthor
Marcelo Dias de Amorim \\
  \affaddr{UPMC Paris Universitas}
}

\newcounter{copyrightbox}

\maketitle

\abstract{
We describe a new model for studying intermittently connected
mobile networks, based on Markovian random temporal graphs,
that captures the influence of message size, maximum tolerated delay
and link stability on the delivery ratio.
}




\section{Introduction}
\label{sec:introduction}
The topology of a real-life network of mobile handheld devices evolves
over time as links come up and down. Successive snapshots of the
evolving connectivity graph yields a \emph{temporal graph}, a
time-indexed sequence of traditional static graphs. These present a
number of new interesting metrics. For example, there might exist a
space-time path between two vertices even if there never exists an end
to end path between them at any given moment. Since such temporal
graphs appear naturally when analyzing connectivity traces in which
nodes periodically scan for neighbors, their theoretical study is
important for understanding the underlying network dynamics.

Modeling temporal networks using random graphs is a relatively
unexplored field. Simple sequences of independent regular random
graphs are used in~\cite{chaintreau_diam} to analyze the diameter of
opportunistic mobile networks. The notion of \emph{connectivity over
  time} is explored in~\cite{pellegrini07} but looses any information
about the order in which contact opportunities appear. 

In this paper, we improve upon previous work, by capturing the strong
real-life correlation between the connectivity graphs at times $t$ and
$t+1$. Since we will be examining how the message size influences the
delivery ratio, it is crucial to model link stability, i.e., the
probability that a link up at time $t$ remains so at time $t+1$. In
order to capture these correlations, we propose a Markovian temporal
graph model.

\section{Model}
\label{sec:model}
\begin{table}[t]
  \centering
  \small
  \caption{Model parameters}
  \begin{tabular}{cl|cl}
  \hline
  $N$ & Number of nodes  &  $d$ & Maximum delay \\
  $\tau$ & Time step &  $r$ & Average link lifetime \\
  $\alpha$ & Packet size &  $\lambda$ & fraction of time a link is down \\
  \hline
  \end{tabular}
  \label{param_desc}
\end{table}

We consider temporal graphs of $N$ mobile nodes that evolve in
discrete time. The time step $\tau$ is equal to the shortest contact
or inter-contact time. In a real-life trace, $\tau$ will be equal to
the sampling period. The only differences between successive time
steps will be which links are up and which are down. They can come up
or go down at the beginning of each time step, but the topology then
remains static until the next time step. 

Each of the potential $\frac{N(N-1)}{2}$ links is considered
independent and is modeled as a two-state ( $\uparrow$ or
$\downarrow$) Markov chain. 
 The evolution of the entire connectivity
graph can also be described as a Markov chain on the tensor product of
the state spaces of all links. 

We note $r$ the average number of time steps that a link spends in the
$\uparrow$ state, and $\lambda$ the fraction of time that a link
spends in the $\downarrow$ state. In a sense, $r$ measures the
evolution speed of the network's topology while $\lambda$ is related
to its density. The average link lifetime is by definition $r \tau$
while the average node degree is $\frac{N-1}{1+\lambda}$. 

When up, all links share the same capacity $\phi$ and thus can
transport the same quantity $\phi \tau$ of information during one
time step. We will refer to $\phi \tau$ as the \emph{link
  size}. Message size is equal to $\alpha \phi \tau$ where $\alpha$
can be greater or smaller than $1$. By abuse of language, we will
refer to $\alpha$ as the message size. For example, a message of size
$2$ ($\alpha=2$) will only be able to traverse links that last for
more than $2$ time steps, whereas a message of size $0.5$, will be
able to perform two hops during each time step. The message size thus
defined is numerically proportional to $\tau$.

Small values of $\tau$ mean that the network topology's
characteristic evolution time is short and thus only small amounts
of information may be transmitted over a link during one time
step. Furthermore we suppose than a mobile application can only
tolerate a given delay in message delivery. We note $d$ the maximum
delay beyond which a delivery is considered to have failed. 

\section{Results}
\label{sec:results}
Using this model, we can derive the delivery ratio of a message using
epidemic routing for $\alpha \le 1$, as well as upper and lower bounds
when $\alpha > 1$. To be successful, the delivery has to occur without
exceeding the maximum allowed delay. Epidemic routing is useful for
theoretical purposes, since its delivery ratio is also that of the
optimal single-copy time-space routing protocol.

\begin{figure}[t]
  \centering
  \includegraphics{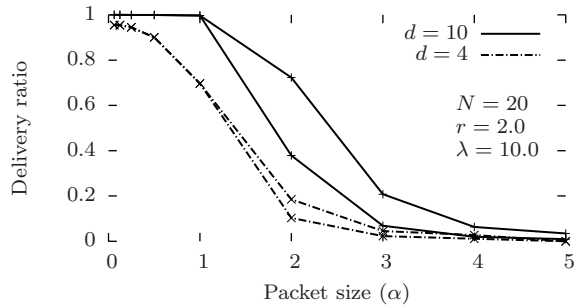}
  \caption{Influence of message size on delivery probability for
    different values of maximum delay ($d$). Each value of $d$
    corresponds to two lines: its upper and lower bounds.}
  \label{param_size}
\end{figure}

\vspace{1mm}
\noindent\textbf{Message size.} (Fig.~\ref{param_size}) Messages
larger than the link size see their delivery probability severely
degraded, though this is somewhat mitigated by longer maximum
delays. On the other hand, messages smaller than the link size are
able of making several hops in a single time step. This is a great
advantage when the time constraints are particularly tight ($d=4$ in
Fig.~\ref{param_size}), but barely has any effect when the time
constraints are looser. This also highlights the influence of node
mobility. Indeed, since the actual message size is proportional to
$\tau$, high node mobility (i.e. small $\tau$) makes the actual link
size smaller and thus further constrains possible message size.

Furthermore, the gain achieved by using small messages is bounded
because because it hits the performance limit of epidemic
routing. Indeed, the best possible epidemic diffusion of a message
will, at each time step, infect a whole connected component if at
least one of its nodes is infected. A small enough packet can spread
sufficiently quickly to achieve this, and thus even smaller packets
bring no performance gain ($d=4$ in Fig.~\ref{param_size}).

\begin{figure}[t]
  \centering
  \subfloat[Number of nodes \label{param_nodes}]{\scalebox{0.8}{\includegraphics{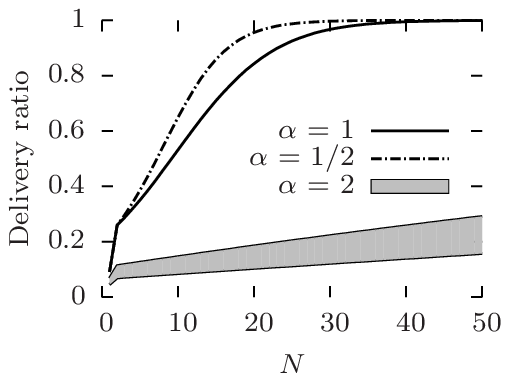}}}
  \subfloat[Average link lifetime \label{param_life}]{\scalebox{0.8}{\includegraphics{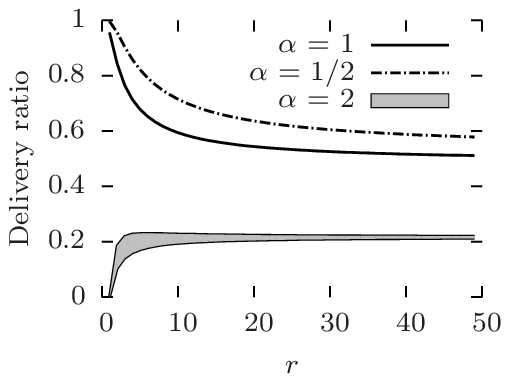}}} \\
  \subfloat[Average node degree \label{param_degree}]{\scalebox{0.8}{\includegraphics{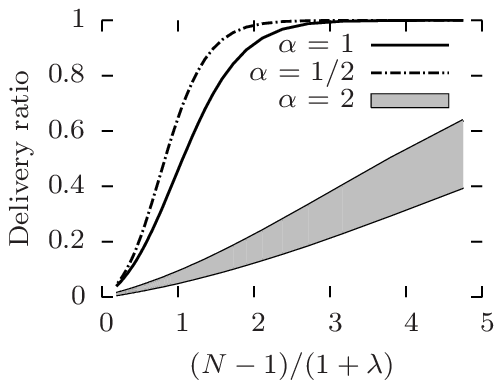}}}
  \subfloat[Maximum delay\label{param_delay}]{\scalebox{0.8}{\includegraphics{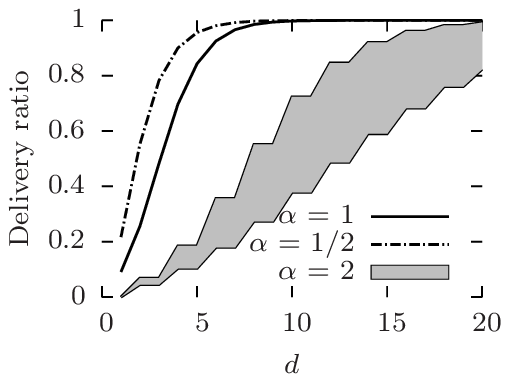}}}
  \caption{Influence of model parameters on the delivery ratio. When unspecified, $N=20$, $r=2$,
    $\lambda=10$ and $d=5$.}
  \label{params_influence}
\end{figure}

\vspace{1mm}
\noindent\textbf{Number of nodes.}
(Fig.~\ref{param_nodes}) The delivery ratio tends to $1$ as $N$
increases. Indeed, for a given source/destination pair, each new node
is a new potential relay in the epidemic dissemination and thus can
only help the delivery ratio.

\vspace{1mm}
\noindent\textbf{Average link lifetime.}
(Fig.~\ref{param_life}) Shorter average link lifetimes make for a more
dynamic network topology. Indeed smaller values of $r$ make for
shorter contact and inter-contact times and increases contact
opportunities. Small messages ($\alpha \le 1$) take advantage of this
and their delivery ratio increases as $r$ decreases. On the other
hand, excessive link instability drives the delivery ratio for larger
messages ($\alpha > 1$) to 0, because fewer links last longer than one
time step.

\vspace{1mm}
\noindent\textbf{Average node degree.}
(Fig.~\ref{param_degree}) Greater connectivity increases the delivery
probability. The sharp slope of the curve when $\alpha \ge 1$ is
reminiscent of percolation in random graphs when the average node
degree hits $1$.

\vspace{1mm}
\noindent\textbf{Maximum Delay.} (Fig.~\ref{param_delay})
All else being equal, there is clearly a threshold value beyond which
almost all messages are delivered. This can be linked to the space-time
diameter of the underlying topology~\cite{chaintreau_diam}.

\vspace{1mm}
\noindent\textbf{Experimental results.} 
We measured the delivery ratio achieved by various message sizes by
replaying a real-life wireless mobility trace
(Rollernet~\cite{tournoux08_rollernet}). Although our model may not be
quantitatively comparable to real-life traces due to unwanted
small-world properties, it accurately predicts the relations between
delivery ratio, maximum delay and message size.

\section{Conclusion}
In this paper, we proposed a new model of random temporal graphs that,
for the first time, captures the correlation between successive
connectivity graph, and provides insights on the interaction between
node mobility, maximum delay and message size. In particular, we have
shown that, given a certain maximum delay and node mobility, message
size has a major impact on the delivery ratio. These results, which we
reproduced by replaying a real-life connectivity
trace, should be taken into consideration
when designing and implementing services for mobile handhelds.

\bibliographystyle{abbrv}
\small
\bibliography{whitbeckmobiheld09}

\begin{thebibliography}{1}

\bibitem{chaintreau_diam}
A.~Chaintreau, A.~Mtibaa, L.~Massoulie, and C.~Diot.
\newblock The diameter of opportunistic mobile networks.
\newblock In {\em CoNEXT '07: Proceedings of the 2007 ACM CoNEXT conference},
  2007.

\bibitem{pellegrini07}
F.~De~Pellegrini, D.~Miorandi, I.~Carreras, and I.~Chlamtac.
\newblock A graph-based model for disconnected ad hoc networks.
\newblock In {\em INFOCOM 2007. 26th IEEE International Conference on Computer
  Communications. IEEE}, 2007.

\bibitem{tournoux08_rollernet}
P.-U. Tournoux, J.~Leguay, F.~Benbadis, V.~Conan, M.~D. de~Amorim, and
  J.~Whitbeck.
\newblock The accordion phenomenon: Analysis, characterization, and impact on
  dtn routing.
\newblock In {\em Proc. IEEE Infocom}, 2009.

\end{thebibliography}

\end{document}